\documentclass{article}
\usepackage{spconf,amsmath,graphicx,hyperref}

\usepackage[utf8]{inputenc} 
\usepackage[T1]{fontenc}    
\usepackage{hyperref}       
\usepackage{url}            
\usepackage{booktabs}       
\usepackage{amsfonts,amssymb}       
\usepackage{nicefrac}       
\usepackage{microtype}      
\usepackage[table]{xcolor}         

\usepackage{amsmath}
\usepackage{amsthm}
\usepackage{graphicx}
\usepackage{adjustbox}
\usepackage{multirow}
\usepackage{wrapfig}
\usepackage{caption}

\usepackage{algorithm}
\usepackage{algpseudocodex}
\usepackage{multirow}
\usepackage{diagbox}
\usepackage{svg}
\usepackage{balance}
\usepackage{makecell}
\usetikzlibrary{arrows.meta}
\usetikzlibrary{shapes.geometric}
\usepackage{threeparttable}

\usepackage{subcaption}
\usepackage[]{caption}
\usetikzlibrary{positioning}

\definecolor{customlinkcolor}{RGB}{155, 155, 255}
\hypersetup{
  colorlinks=true, 
  pdfborder={0 0 0}, 
  linkcolor=black, 
  citecolor=black, 
  urlcolor=black 
}
\usepackage{fontenc}
\usepackage[style=ieee,backend=biber,maxnames=3,minnames=3,citestyle=numeric-comp,eprint=true,date=short  ]{biblatex}
\AtBeginBibliography{%
  \small   
}
\makeatletter
\newcommand\fs@betterruled{%
  \def\@fs@cfont{\bfseries}\let\@fs@capt\floatc@ruled
  \def\@fs@pre{\vspace*{5pt}\hrule height.8pt depth0pt \kern2pt}%
  \def\@fs@post{\kern2pt\hrule\relax}%
  \def\@fs@mid{\kern2pt\hrule\kern2pt}%
  \let\@fs@iftopcapt\iftrue}
\floatstyle{betterruled}
\restylefloat{algorithm}
\makeatother

\newtheorem{remark}{Remark}

\definecolor{bl}{rgb}{0.25, 0.5, 0.9}
\newcommand{\firstres}[1]{{\textbf{#1}}}
\newcommand{\secondres}[1]{{\underline{#1}}}

\def\Xin{X^\mathrm{in}}
\def\Xtarget{X^\mathrm{target}}
\def\Xout{\hat{X}^\mathrm{out}}
\def\Wone{I_1}
\def\Wtwo{I_2}
\def\Wthree{I_3}

\usepackage{marvosym}
\usepackage{pifont}
\usepackage{xspace}
\newcommand{\OurModel}{\textsc{CTPNet}\xspace}

\usepackage[capitalize]{cleveref}

\Crefname{figure}{Fig.}{Figs.}

\Crefname{definition}{Definition}{Figs.}

\Crefformat{equation}{Eq.~(#1)}

\crefname{appendix}{App.}{Apps.}
\Crefname{appendix}{Appendix}{Appendixs}
\Crefname{algorithm}{Alg.}{Algs.}
\usepackage{fancyhdr} 

\title{Channel, Trend and Periodic-Wise Representation Learning for Multivariate Long-term Time Series Forecasting}

\name{
\hspace{-1.6em}
Zhangyao Song\textsuperscript{*}, ~Nanqing Jiang\textsuperscript{*},~Miaohong He,~Xiaoyu Zhao,~Tao Guo\textsuperscript{\Letter} 
\thanks{This work was supported in part by the NSFC Project 62301144, and in part by the Zhishan Young Scholar Fund No.2242025RCB0032.  
The first two authors are equally contributed. 
(Corresponding author: Tao Guo, taoguo@seu.edu.cn).
\href{https://github.com/jiangnanqing/CTPNet.git}{\color{red}Code}
}}
\address{
School of Cyber Science and Engineering, Southeast University, Nanjing, China}

\fancypagestyle{firstpage}{
    \fancyhf{} 
    \fancyhead[L]{Published as a conference paper in ICASSP 2026} 
}
\begin{document}
\ninept
\maketitle
\begin{abstract}
Downsampling-based methods for time series forecasting have gained significant attention due to their effectiveness in capturing sequence trends. However, these approaches primarily focus on dependencies within subsequences, neglecting inter-subsequence and inter-channel interactions, which limits forecasting accuracy. To overcome these limitations, we propose \OurModel, a novel framework that explicitly learns representations from three complementary perspectives: 
i) inter-channel dependencies, captured by a temporal query-based multi-head attention mechanism; 
ii) intra-subsequence dependencies, modeled via a Transformer to characterize trend variations; and 
iii) inter-subsequence dependencies, extracted by reusing the encoder with residual connections to capture global periodic patterns. By jointly integrating these levels, the proposed method provides a holistic representation of temporal dynamics. Extensive experiments on multiple real-world datasets demonstrate the superiority and robustness of \OurModel, establishing new state-of-the-art performance in long-term time series forecasting.
\end{abstract}

\section{Introduction}

Time series forecasting aims to predict future values based on historical time series data and has significant applications in fields such as energy management~\cite{zhou-2022-fedformer}, weather forecasting~\cite{ilbert-2024-samformer}, and traffic monitoring~\cite{woo-2022-etsformer,zhou-2021-informer}. Furthermore, multivariate long-term time series forecasting has garnered increasing attention due to its ability to provide richer future insights and leverage dependencies between multiple variables~\cite{lin-2025-TQNet,zhang-2023-crossformer,zhang-2025-channelmixer}. With advancements in deep learning, various methods have been proposed for multivariate long-term time series forecasting, including those based on Recurrent Neural Networks (RNNs)~\cite{lin-2023-segrnn,zhang-2023-rnn_nc}, Convolutional Neural Networks (CNNs)~\cite{luo-2024-moderntcn,zhang-2025-channelmixer,bai-2028-tcn,liu-2022-scinet}, Multi-layer Perceptrons (MLPs)~\cite{zeng-2023-dlinear,lin-2024-cyclenet}, and Transformers~\cite{zhou-2021-informer,liu-2024-itransformer,nie-2023-patchtst}. However, the complexity of real-world time series data often limits the effectiveness of these methods.

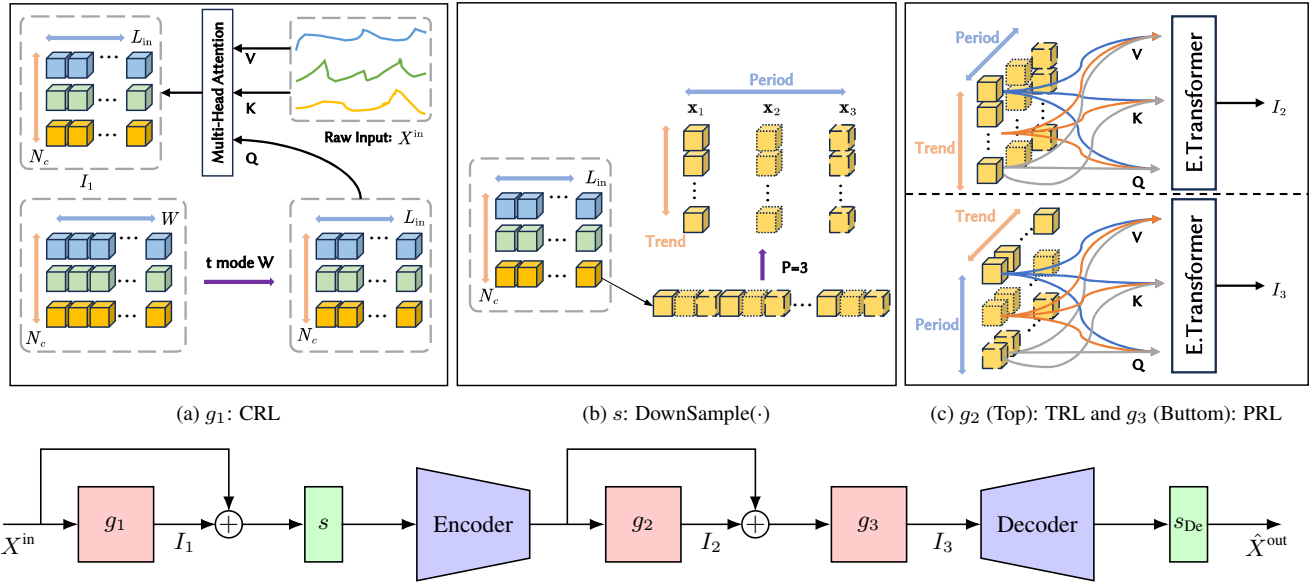
\begin{figure*}[t]
    \centering

    \begin{subfigure}[t]{0.33\linewidth}
        \centering
        \includegraphics[width=\linewidth]{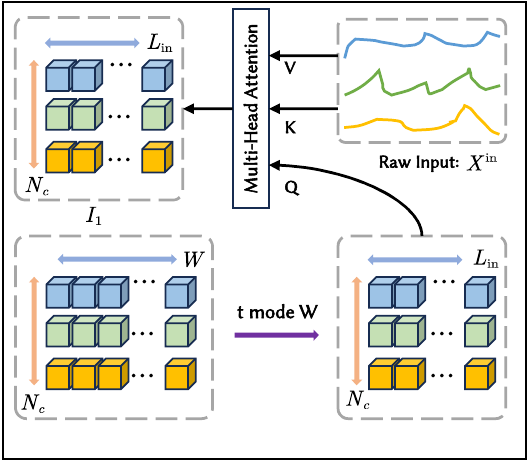}
        \caption{$g_1$: CRL}
        \label{fig:g1}
    \end{subfigure}
    \begin{subfigure}[t]{0.33\linewidth}
        \centering
        \includegraphics[width=\linewidth]{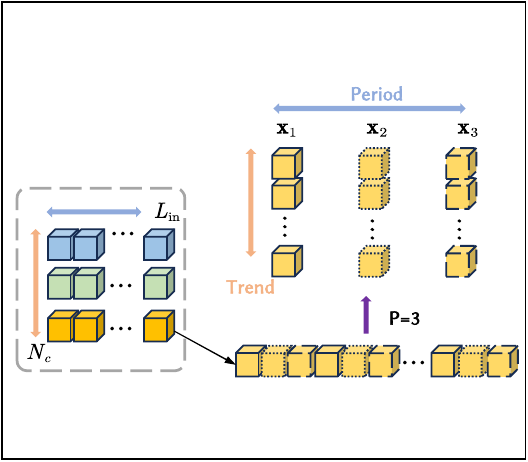}
        \caption{$s$: DownSample($\cdot$)}
        \label{fig:downsample}
    \end{subfigure}
    \begin{subfigure}[t]{0.3035\linewidth}
        \centering
        \includegraphics[width=\linewidth]{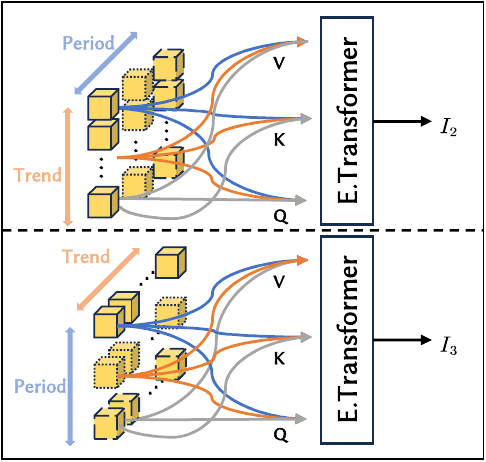}
        \caption{$g_2$ (Top): TRL and $g_3$ (Buttom): PRL}
        \label{fig:g3}
    \end{subfigure}
    \vspace{7pt}

    \begin{subfigure}[t]{\linewidth}
        \begin{tikzpicture}[>={Latex},line width=0.6pt]
            \draw[->] (0,0) -- (1,0) node[pos=0.2,below]{$\Xin$};
            \draw[fill=red!20] (1,-0.5) -- (1,0.5) -- (2,0.5) -- (2,-0.5) -- cycle;
            \node at (1.5,0) {$g_1$};
            \draw[->] (2,0) -- (2.83,0) node[pos=0.5,below]{$\Wone$};
            \draw[->] (0.5,0) -- (0.5,1) -- (3,1) -- (3,0.17);
            \node[draw,circle,minimum size=0.1cm,inner sep=0pt] at (3,0) {\fontsize{10}{10}\selectfont$+$};

            \draw[->] (3.17,0) -- (4,0);

            \node[fill=green!20,draw, minimum width=0.5cm, minimum height=1cm, inner sep=0pt, anchor=west] at (4,0) {$s$};

            \draw[->] (4.5,0) -- (5.5,0);

            \draw[fill=blue!20] (5.5,-0.75) -- (5.5,0.75) -- (7,0.45) -- (7,-0.45) -- cycle;
            \node at (6.25,0) {Encoder};

            \draw[->] (7,0) -- (8,0);
            \node[draw, minimum width=1cm, minimum height=1cm, inner sep=0pt, anchor=west,fill=red!20] at (8,0) {$g_2$};
            \draw[->] (9,0) -- (9.83,0) node[pos=0.5,below]{$\Wtwo$};
            \draw[->] (7.5,0) -- (7.5,1) -- (10,1) -- (10,0.17);
            \node[draw,circle,minimum size=0.1cm,inner sep=0pt] at (10,0) {\fontsize{10}{10}\selectfont$+$};
            \draw[->] (10.17,0) -- (11,0);
            \node[draw, minimum width=1cm, minimum height=1cm, inner sep=0pt, anchor=west,fill=red!20] at (11,0) {$g_3$};
            \draw[->] (12,0) -- (13,0) node[pos=0.5,below]{$\Wthree$};

            \draw[fill=blue!20] (13,-0.45) -- (13,0.45) -- (14.5,0.75) -- (14.5,-0.75) -- cycle;
            \node at (13.75,0) {Decoder};
            \draw[->] (14.5,0) -- (15.5,0);
            \node[fill=green!20,draw, minimum width=0.5cm, minimum height=1cm, inner sep=0pt, anchor=west] at (15.5,0) {$s_\text{De}$};
            \draw[->] (16,0) -- (17,0) node[pos=0.8,below]{$\Xout$};
        \end{tikzpicture}
    \end{subfigure}
    \caption{\textbf{Overview of \OurModel}.
     Here $g_1$, $g_2$ and $g_3$ represent the representation learning networks for inter-channel dependencies, intra-subsequence dependencies, and inter-subsequence dependencies, respectively. 
    $s$ denotes $\text{Downsample}(\cdot)$, and $s_\text{De}$ denotes it's inverse process.
    }
    \label{fig:overview}
    \vspace{-12pt}

\end{figure*}

Accurate long-term time series forecasting fundamentally relies on capturing the inherent inter-channel dependencies and temporal dependencies within the data. Temporal dependencies can be further categorized into trends and periodic patterns. For inter-channel dependencies, recent studies~\cite{zhang-2023-crossformer,lin-2025-TQNet,liu-2024-itransformer} have utilized attention mechanisms due to their strong capability to extract multi-level representations. For temporal dependencies, a common strategy~\cite{lin-2024-sparsetsf,wu-2023-timesnet,wang-2025-timemixerplus} is to downsample (see~\Cref{fig:downsample}) the time series into subsequences based on periodicity, which has demonstrated strong predictive capabilities. At this stage, each subsequence captures the trend variations of the time series, while relationships across subsequences reflect periodic dependencies. For instance, \cite{lin-2024-sparsetsf} employs a weight-shared MLP to extract intra-subsequence dependencies. Since time series often exhibit fixed periodic patterns (e.g., daily cycles), each downsampled subsequence tends to display consistent trends. However, this approach primarily focuses on dependencies within individual subsequences, neglecting inter-subsequence dependencies, which leads to the loss of valuable temporal information. Moreover, it overlooks the modeling of inter-channel dependencies~\cite{yu-2024-Leddam}. These limitations raise a fundamental question: how can we design a forecasting framework that effectively learns comprehensive temporal representations across inter-channel, intra-subsequence, and inter-subsequence dependencies?

In this paper, we address these limitations by proposing \OurModel (\textbf{C}hannel, \textbf{T}rend, and \textbf{P}eriod-wise representation learning \textbf{Net}work), a novel framework that performs representation learning on three complementary levels of dependencies: i) $\Wone$ inter-channel dependencies, which capture interactions across variables; ii) $\Wtwo$ intra-subsequence dependencies, which model local temporal structures within each subsequence; and iii) $\Wthree$ inter-subsequence dependencies, which uncover global temporal dynamics across subsequences. Specifically, for inter-channel dependencies, we employ a temporal query-based~\cite{lin-2025-TQNet} Multi-Head Attention (MHA)~\cite{vaswani-2017-attentionisallyouneed}. By using a learnable query matrix $Q$, the MHA mechanism effectively captures cross-channel dependencies. For intra-subsequence dependencies, we adopt a Transformer with linear complexity~\cite{shen-2019-efficient}. The residual connection~\cite{he-2016-resnet} between the inter-channel dependencies and the original input, followed by linear encoding~\cite{zeng-2023-dlinear}, is used as the Query, Key, and Value of the attention mechanism. This enables the Transformer to effectively capture complex trend variations (i.e., intra-subsequence dependencies) in the high-dimensional hidden space. Finally, for inter-subsequence dependencies, we use the same network as before. By inputting the transposed encoder output with a residual connection to $\Wtwo$, the network can model inter-subsequence dependencies.

Using these three perspectives together, \OurModel offers a holistic representation of time series data, overcoming the limitations of previous methods. Extensive experiments conducted on multiple benchmark datasets demonstrate the superiority of our approach. In summary, our contributions are threefold:
\begin{itemize}
    \item Unlike previous methods that only exploit intra-subsequence dependencies within downsampled subsequences, we explicitly model inter-subsequence dependencies.
    \item We propose \OurModel, a unified framework that simultaneously learns inter-channel, intra-subsequence, and inter-subsequence dependencies for comprehensive representation.
    \item Through extensive experiments, we establish new state-of-the-art performance across multiple benchmarks, demonstrating the robustness and generality of our framework.
\end{itemize}



\section{Preliminaries}\label{sec:preliminaries}

\subsection{Problem Statement}
Given a historical series $\Xin =  [x_{t}^{(1:N_c)},\cdots, x_{t-L_\mathrm{in}-1}^{(1:N_c)}]\in \mathbb{R}^{N_c\times  L_\mathrm{in}}$ of $L_\mathrm{in}$ time steps with $N_c$ channels (e.g., the number of variables), the goal of time series forecasting (TSF) is to predict the series $\Xout=  [\hat{x}_{t+L_\mathrm{out}}^{(1:N_c)},\cdots, \hat{x}_{t+1}^{(1:N_c)}] \in \mathbb{R}^{N_c\times  L_\mathrm{out}}$ for the next $L_\mathrm{out}$ time steps based on $\Xin$, while minimizing the difference between $\Xout$ and the target $\Xtarget$. This can be formulated as:
\begin{equation}
\label{eq:tsf}
 [\hat{x}_{t+L_\mathrm{out}}^{(1:N_c)},\cdots, \hat{x}_{t+1}^{(1:N_c)}]=f \left([x_{t}^{(1:N_c)},\cdots, x_{t-L_\mathrm{in}-1}^{(1:N_c)}]\right),
\end{equation}
where $f$ represents a trainable neural network. 

\subsection{Time Series Downsampling}\label{sec:downsample}

Given a time series $X=[x_{t}^{(n)},\cdots, x_{t-L_\mathrm{in}-1}^{(n)}]$ with periodicity $P$, the operation of downsampling~\cite{lin-2024-sparsetsf} decomposes the sequence into $P$ downsampled subsequences of length $N_\mathrm{pin}=\lfloor \frac{L_\mathrm{in}}{P} \rfloor$. As illustrated in \Cref{fig:downsample}, this process can be expressed as:
\begin{equation}
    [\mathbf{x}_{1},\cdots, \mathbf{x}_{P}]  =\mathrm{DownSample}(X),
\end{equation}
where $\mathbf{x}_{p}=[x_{t+p-1+0\times P},\cdots,x_{t+p-1+(N_\mathrm{pin}-1)\times P}]\in\mathbb{R}^{N_\mathrm{pin}}$ denotes the $p$-th subsequence. The inverse process of $\mathrm{DownSample}(\cdot)$ is denoted as $\mathrm{De{\text -}DownSample}(\cdot)$.

\section{Method: CTPNET}\label{sec:method}
\renewcommand{\arraystretch}{0.85}
\begin{table*}[t]
    \vspace{-12pt}
    \centering
    \caption{Multivariate time series forecasting results
    }\label{tab:full_results}
    \footnotesize
    \vspace{-6pt}

    \setlength{\tabcolsep}{4pt}

    \begin{adjustbox}{max width=\linewidth}
        \begin{threeparttable}
            {\sc
                \begin{tabular}{@{}rccccccccccccccccc@{}}
                    \toprule
                    \multicolumn{2}{c}{\multirow{2}{*}{Model}} & \multicolumn{2}{c}{{\OurModel}} & \multicolumn{2}{c}{{TQNet~\cite{lin-2025-TQNet}}} & \multicolumn{2}{c}{{Twins*~\cite{stitsyuk-2025-xpatch}}} & \multicolumn{2}{c}{{Leddam~\cite{yu-2024-Leddam}}} & \multicolumn{2}{c}{{iTrans*~\cite{liu-2024-itransformer}}} & \multicolumn{2}{c}{{PatchTST~\cite{nie-2023-patchtst}}} & \multicolumn{2}{c}{{Cross*~\cite{zhang-2023-crossformer}}} & \multicolumn{2}{c}{{FED*~\cite{zhou-2022-fedformer}}}                                                                                                                                                                                  \\

                    \multicolumn{1}{c}{}                       & \multicolumn{1}{c}{}            & \multicolumn{2}{c}{(Ours)}                        & \multicolumn{2}{c}{(2025)}                               & \multicolumn{2}{c}{(2025)}                         & \multicolumn{2}{c}{(2024)}                                 & \multicolumn{2}{c}{(2024)}                              & \multicolumn{2}{c}{(2023)}                                 & \multicolumn{2}{c}{(2023)}                            & \multicolumn{2}{c}{(2022)}                                                                                                                                                     \\

                    \multicolumn{2}{c}{Metric}                 & MSE                             & \multicolumn{1}{c}{MAE}                           & MSE                                                      & \multicolumn{1}{c}{MAE}                            & MSE                                                        & \multicolumn{1}{c}{MAE}                                 & MSE                                                        & \multicolumn{1}{c}{MAE}                               & MSE                        & \multicolumn{1}{c}{MAE} & MSE   & \multicolumn{1}{c}{MAE} & MSE               & \multicolumn{1}{c}{MAE} & MSE   & \multicolumn{1}{c}{MAE}         \\
                    \midrule
                    \multirow{4}{*}{\rotatebox{90}{ETTh1}}     & 96                              & \firstres{0.363}                                  & \firstres{0.388}                                         & \secondres{0.371}                                  & \secondres{0.393}                                          & 0.385                                                   & 0.401                                                      & 0.377                                                 & 0.394                      & 0.386                   & 0.405 & 0.414                   & 0.419             & 0.423                   & 0.448 & 0.376                   & 0.419 \\
                                                               & 192                             & \firstres{0.414}                                  & \firstres{0.416}                                         & 0.428                                              & 0.426                                                      & 0.439                                                   & 0.431                                                      & 0.424                                                 & \secondres{0.422}          & 0.441                   & 0.436 & 0.460                   & 0.445             & 0.471                   & 0.474 & \secondres{0.420}       & 0.448 \\
                                                               & 336                             & \firstres{0.453}                                  & \firstres{0.435}                                         & 0.476                                              & 0.446                                                      & 0.480                                                   & 0.452                                                      & \secondres{0.459}                                     & \secondres{0.442}          & 0.487                   & 0.458 & 0.501                   & 0.466             & 0.570                   & 0.546 & \secondres{0.459}       & 0.465 \\
                                                               & 720                             & \firstres{0.451}                                  & \firstres{0.451}                                         & 0.487                                              & 0.470                                                      & 0.480                                                   & 0.474                                                      & \secondres{0.463}                                     & \secondres{0.459}          & 0.503                   & 0.491 & 0.500                   & 0.488             & 0.653                   & 0.621 & 0.506                   & 0.507 \\
                    \hline
                    \multirow{4}{*}{\rotatebox{90}{ETTh2}}     & 96                              & \firstres{0.287}                                  & \firstres{0.336}                                         & 0.295                                              & \secondres{0.343}                                          & \secondres{0.292}                                       & 0.345                                                      & \secondres{0.292}                                     & \secondres{0.343}          & 0.297                   & 0.349 & 0.302                   & 0.348             & 0.745                   & 0.584 & 0.358                   & 0.397 \\
                                                               & 192                             & \firstres{0.356}                                  & \firstres{0.381}                                         & \secondres{0.367}                                  & 0.393                                                      & 0.375                                                   & 0.395                                                      & \secondres{0.367}                                     & \secondres{0.389}          & 0.380                   & 0.400 & 0.388                   & 0.400             & 0.877                   & 0.656 & 0.429                   & 0.439 \\
                                                               & 336                             & \firstres{0.412}                                  & \firstres{0.422}                                         & 0.417                                              & 0.427                                                      & 0.417                                                   & 0.429                                                      & \firstres{0.412}                                      & \secondres{0.424}          & 0.428                   & 0.432 & 0.426                   & 0.433             & 1.043                   & 0.731 & 0.496                   & 0.487 \\
                                                               & 720                             & \secondres{0.410}                                 & \secondres{0.433}                                        & 0.433                                              & 0.446                                                      & \firstres{0.406}                                        & \firstres{0.430}                                           & 0.419                                                 & 0.438                      & 0.427                   & 0.445 & 0.431                   & 0.446             & 0.463                   & 1.104 & 0.763                   & 0.474 \\
                    \hline
                    \multirow{4}{*}{\rotatebox{90}{ETTm1}}     & 96                              & \firstres{0.310}                                  & \firstres{0.346}                                         & \secondres{0.311}                                  & \secondres{0.353}                                          & 0.325                                                   & 0.364                                                      & 0.319                                                 & 0.359                      & 0.334                   & 0.368 & 0.329                   & 0.367             & 0.404                   & 0.426 & 0.379                   & 0.419 \\
                                                               & 192                             & \secondres{0.359}                                 & \firstres{0.376}                                         & \firstres{0.356}                                   & \secondres{0.378}                                          & 0.372                                                   & 0.390                                                      & 0.369                                                 & 0.383                      & 0.377                   & 0.391 & 0.367                   & 0.385             & 0.450                   & 0.451 & 0.426                   & 0.441 \\
                                                               & 336                             & \firstres{0.387}                                  & \firstres{0.395}                                         & \secondres{0.390}                                  & \secondres{0.401}                                          & 0.406                                                   & 0.412                                                      & 0.394                                                 & 0.402                      & 0.426                   & 0.420 & 0.399                   & 0.410             & 0.532                   & 0.515 & 0.445                   & 0.459 \\
                                                               & 720                             & \firstres{0.449}                                  & \firstres{0.434}                                         & \secondres{0.452}                                  & 0.440                                                      & 0.467                                                   & 0.448                                                      & 0.460                                                 & 0.442                      & 0.491                   & 0.459 & 0.454                   & \secondres{0.439} & 0.666                   & 0.589 & 0.543                   & 0.490 \\
                    \hline
                    \multirow{4}{*}{\rotatebox{90}{ETTm2}}     & 96                              & 0.174                                             & \firstres{0.251}                                         & \firstres{0.173}                                   & \secondres{0.256}                                          & \firstres{0.173}                                        & \secondres{0.256}                                          & 0.176                                                 & 0.257                      & 0.180                   & 0.264 & 0.175                   & 0.259             & 0.287                   & 0.366 & 0.203                   & 0.287 \\
                                                               & 192                             & \firstres{0.238}                                  & \firstres{0.294}                                         & \firstres{0.238}                                   & \secondres{0.298}                                          & 0.239                                                   & 0.300                                                      & 0.243                                                 & 0.303                      & 0.250                   & 0.309 & 0.241                   & 0.302             & 0.414                   & 0.492 & 0.269                   & 0.328 \\
                                                               & 336                             & \secondres{0.299}                                 & \firstres{0.333}                                         & 0.301                                              & 0.340                                                      & \firstres{0.298}                                        & \secondres{0.339}                                          & 0.303                                                 & 0.341                      & 0.311                   & 0.348 & 0.305                   & 0.343             & 0.597                   & 0.542 & 0.325                   & 0.366 \\
                                                               & 720                             & 0.399                                             & \firstres{0.392}                                         & \firstres{0.397}                                   & \secondres{0.396}                                          & \firstres{0.397}                                        & 0.397                                                      & 0.400                                                 & 0.398                      & 0.412                   & 0.407 & 0.402                   & 0.400             & 1.730                   & 1.042 & 0.421                   & 0.415 \\
                    \hline

                    \multirow{4}{*}{\rotatebox{90}{ECL}}       & 96                              & \firstres{0.134}                                  & \firstres{0.224}                                         & \firstres{0.134}                                   & \secondres{0.229}                                          & 0.139                                                   & 0.233                                                      & 0.141                                                 & 0.235                      & 0.148                   & 0.240 & 0.181                   & 0.270             & 0.219                   & 0.314 & 0.193                   & 0.308 \\
                                                               & 192                             & \firstres{0.151}                                  & \firstres{0.240}                                         & \secondres{0.154}                                  & \secondres{0.247}                                          & 0.158                                                   & 0.252                                                      & 0.159                                                 & 0.252                      & 0.162                   & 0.253 & 0.188                   & 0.274             & 0.231                   & 0.322 & 0.201                   & 0.315 \\
                                                               & 336                             & \firstres{0.166}                                  & \firstres{0.256}                                         & \secondres{0.169}                                  & \secondres{0.264}                                          & 0.172                                                   & 0.267                                                      & 0.173                                                 & 0.268                      & 0.178                   & 0.269 & 0.204                   & 0.293             & 0.246                   & 0.337 & 0.214                   & 0.329 \\
                                                               & 720                             & \firstres{0.200}                                  & \firstres{0.286}                                         & 0.201                                              & 0.294                                                      & \firstres{0.200}                                        & \secondres{0.293}                                          & 0.201                                                 & 0.295                      & 0.225                   & 0.317 & 0.246                   & 0.324             & 0.280                   & 0.363 & 0.246                   & 0.355 \\
                    \hline
                    \multirow{4}{*}{\rotatebox{90}{Solar}}     & 96                              & \firstres{0.168}                                  & \firstres{0.223}                                         & \secondres{0.173}                                  & 0.233                                                      & 0.193                                                   & \secondres{0.224}                                          & 0.197                                                 & 0.241                      & 0.203                   & 0.237 & 0.234                   & 0.286             & 0.310                   & 0.331 & 0.242                   & 0.342 \\
                                                               & 192                             & \firstres{0.184}                                  & \firstres{0.238}                                         & \secondres{0.199}                                  & 0.257                                                      & 0.223                                                   & \secondres{0.250}                                          & 0.231                                                 & 0.264                      & 0.233                   & 0.261 & 0.267                   & 0.310             & 0.734                   & 0.725 & 0.285                   & 0.380 \\
                                                               & 336                             & \firstres{0.200}                                  & \firstres{0.254}                                         & \secondres{0.211}                                  & \secondres{0.263}                                          & 0.246                                                   & 0.268                                                      & 0.241                                                 & 0.268                      & 0.248                   & 0.273 & 0.290                   & 0.315             & 0.750                   & 0.735 & 0.282                   & 0.376 \\
                                                               & 720                             & \firstres{0.204}                                  & \firstres{0.247}                                         & \secondres{0.209}                                  & \secondres{0.270}                                          & 0.245                                                   & 0.272                                                      & 0.250                                                 & 0.281                      & 0.249                   & 0.275 & 0.289                   & 0.317             & 0.769                   & 0.765 & 0.357                   & 0.427 \\
                    \hline
                    \multirow{4}{*}{\rotatebox{90}{Traffic}}   & 96                              & 0.398                                             & \firstres{0.251}                                         & 0.413                                              & 0.261                                                      & \firstres{0.382}                                        & 0.260                                                      & 0.426                                                 & 0.276                      & \secondres{0.395}       & 0.268 & 0.462                   & 0.290             & 0.522                   & 0.290 & 0.587                   & 0.366 \\
                                                               & 192                             & 0.419                                             & \firstres{0.261}                                         & 0.432                                              & 0.271                                                      & \firstres{0.392}                                        & \secondres{0.267}                                          & 0.458                                                 & 0.289                      & \secondres{0.417}       & 0.276 & 0.466                   & 0.290             & 0.530                   & 0.293 & 0.604                   & 0.373 \\
                                                               & 336                             & 0.436                                             & \firstres{0.270}                                         & 0.450                                              & 0.277                                                      & \firstres{0.410}                                        & \secondres{0.276}                                          & 0.486                                                 & 0.297                      & \secondres{0.433}       & 0.283 & 0.482                   & 0.300             & 0.558                   & 0.305 & 0.621                   & 0.383 \\
                                                               & 720                             & 0.470                                             & \firstres{0.291}                                         & 0.486                                              & 0.295                                                      & \firstres{0.442}                                        & \secondres{0.292}                                          & 0.498                                                 & 0.313                      & \secondres{0.467}       & 0.302 & 0.514                   & 0.320             & 0.589                   & 0.328 & 0.626                   & 0.382 \\
                    \hline
                    \multirow{4}{*}{\rotatebox{90}{Weather}}   & 96                              & \firstres{0.156}                                  & \firstres{0.196}                                         & 0.157                                              & \secondres{0.200}                                          & 0.161                                                   & 0.201                                                      & \firstres{0.156}                                      & 0.202                      & 0.174                   & 0.214 & 0.177                   & 0.210             & 0.158                   & 0.230 & 0.217                   & 0.296 \\
                                                               & 192                             & \firstres{0.204}                                  & \firstres{0.241}                                         & \secondres{0.206}                                  & \secondres{0.245}                                          & 0.211                                                   & 0.248                                                      & 0.207                                                 & 0.250                      & 0.221                   & 0.254 & 0.225                   & 0.250             & \secondres{0.206}       & 0.277 & 0.276                   & 0.336 \\
                                                               & 336                             & \firstres{0.262}                                  & \firstres{0.285}                                         & \firstres{0.262}                                   & \secondres{0.287}                                          & 0.266                                                   & 0.291                                                      & \firstres{0.262}                                      & 0.291                      & 0.278                   & 0.296 & 0.278                   & 0.290             & 0.272                   & 0.335 & 0.339                   & 0.380 \\
                                                               & 720                             & \firstres{0.343}                                  & \firstres{0.340}                                         & 0.344                                              & 0.342                                                      & 0.347                                                   & 0.343                                                      & \firstres{0.343}                                      & 0.343                      & 0.358                   & 0.349 & 0.354                   & \firstres{0.340}  & 0.398                   & 0.418 & 0.403                   & 0.428 \\
                    \hline
                    \multicolumn{2}{c|}{$1^\text{st}$ Count}   & \multicolumn{2}{c}{54}          & \multicolumn{2}{c}{6}                             & \multicolumn{2}{c}{10}                                   & \multicolumn{2}{c}{4}                              & \multicolumn{2}{c}{0}                                      & \multicolumn{2}{c}{1}                                   & \multicolumn{2}{c}{0}                                      & \multicolumn{2}{c}{0}                                                                                                                                                                                                                  \\
                    \bottomrule
                \end{tabular}}
            \begin{tablenotes}
                \small
                \item The look-back length $L_\text{in}$ is fixed at 96, and "*" denotes "former".
            \end{tablenotes}
        \end{threeparttable}
    \end{adjustbox}
    \vspace{-12pt}
\end{table*}

\subsection{Structure Overview}\label{sec:overview}
As illustrated in \Cref{fig:overview}, the proposed \OurModel consists of seven key components:
\ding{172} \textbf{Channel-wise Representation Learning} (CRL) captures inter-channel dependencies;
\ding{173} $\mathrm{Downsample}(\cdot)$ downsamples the residual connection of $\Wone$ and the original input to generate multiple subsequences;
\ding{174} \textbf{Encoder} maps each subsequence of length $N_\mathrm{pin}$ into a $D$-dimensional representation;
\ding{175} \textbf{Trend-wise Representation Learning} (TRL) captures intra-subsequence dependencies;
\ding{176} \textbf{Periodic-wise Representation Learning} (PRL) captures inter-subsequence dependencies;
\ding{177} \textbf{Decoder} converts the subsequence length from $D$ back to $N_\mathrm{pout}$, enabling the next operation to yield a prediction horizon of length $L_\mathrm{out}$;
\ding{178} $\mathrm{De-Downsample}(\cdot)$ recombines the multiple subsequences into a single sequence.

\subsection{Channel-wise Representation Learning (CRL)}
We adopt the Temporal Query (TQ)~\cite{lin-2025-TQNet} technique to model inter-channel dependencies $\Wone$. As shown in \cref{fig:g1}, this method introduces periodically shifted learnable vectors as queries to mitigate the impact of noise and missing values on local correlations, while the raw input sequence serves as keys and values to preserve instance-level information. This allows the attention mechanism to integrate both global priors and local features. Specifically, given a period length $W$, we define the learnable parameters
\begin{equation}
    \theta^{TQ} \in \mathbb{R}^{N_c\times W},
\end{equation}
and at time step $t$, a segment of length $L_\mathrm{in}$ is selected as the query:
\begin{equation}
    Q = \theta^{TQ}_{t,L_\mathrm{in}} \in \mathbb{R}^{N_c \times L_\mathrm{in}},
\end{equation}
where the index is determined by $t \bmod W$, enabling periodic parameter reuse. The input sequence $X_t$ is linearly projected to obtain keys and values:
\begin{equation}
    K_1 = X_t W^{K_1}, \quad V_1 = X_t W^{V_1},
\end{equation}
and the attention is computed as:
\begin{equation}
    \mathrm{Head}_h = \mathrm{Softmax}\!\left(\frac{Q_h K_h^\top}{\sqrt{L_\mathrm{in}}}\right)V_h.
\end{equation}
The MHA mechanism concatenates the outputs:
\begin{equation}
    \Wone=\mathrm{ MHA }(Q, K_1, V_1)=\mathrm{Concat}\left(\mathrm{head}_{1}, \ldots, \mathrm{head}_{H}\right) W^{O},
    \label{eq:w1}
\end{equation}
where $W^{O}$ is a learnable projection matrix.
\begin{remark}
    It is worth noting that the period $W$ here differs from the one used in downsampling. For downsampling, a large period would result in overly sparse sequences and loss of temporal information; thus, the smallest inherent period (e.g., daily period of 24) is usually adopted. In contrast, in the TQ mechanism, the query $Q$ is designed to capture long-range periodic variations, and therefore a longer period (e.g., weekly period of 168) is preferred. Both types of periods can be determined via the autocorrelation function (ACF)~\cite{madsen-2007-acf}.
\end{remark}
We perform a residual connection between $\Wone$ and the original input $\Xin$, formulated as:
\begin{equation}
    {\Xin}'\in \mathbb{R}^{N_c\times  L_\mathrm{in}} =\Xin+ \Wone,
\end{equation}
to enhance training stability. After that, downsampling (as introduced in \cref{sec:downsample}) is applied:
\begin{equation}
    {\Xin_\mathrm{ds}}'\in \mathbb{R}^{N_c\times P \times N_\mathrm{pin}}=\mathrm{DownSample}({\Xin}'),
\end{equation}
and the resulting $P$ subsequences are fed into the following networks.

\subsection{Trend-wise Representation Learning (TRL)}
The downsampled subsequences are processed by a linear encoder, which maps the temporal dimension after downsampling into a $D$-dimensional hidden space:
\begin{equation}
    Z \in \mathbb{R}^{N_c \times P \times D} = \mathrm{Encoder}({\Xin_\mathrm{ds}}').
\end{equation}
Prior studies~\cite{lin-2024-sparsetsf,lin-2024-cyclenet,zeng-2023-dlinear} have shown that this setup allows neural networks to extract richer features in the higher-dimensional hidden space while alleviating the sparsity issue caused by wide intervals in downsampling. Afterward, the TRL network captures intra-subsequence temporal dependencies $\Wtwo$ within the hidden space, as illustrated in \cref{fig:g3} Top. Specifically, we adopt an efficient attention~\cite{shen-2019-efficient} to extract $\Wtwo$, defined as:
\begin{equation}
    E_2=\mathrm{Attention}(Q_2, K_2, V_2) = \rho(Q_2)\left(\rho(K_2^\top)\cdot V_2\right),
    \label{eq:e2}
\end{equation}
where $\rho(\cdot)$ denotes $\frac{\cdot}{\sqrt{\dim(\cdot)}}$ and:
\begin{equation}
    Q_2 = Z W^{Q_2}, \quad K_2 = Z W^{K_2}, \quad V_2 = Z W^{V_2},
    \label{eq:qkv2}
\end{equation}
with $W^{Q_2}, W^{K_2}, W^{V_2}$ being learnable weight matrices. $E_2$ is input into an MHA, following an operation similar to that in \cref{eq:w2}, to obtain $E_2^\mathrm{(mha)}$. Then, $E_2^\mathrm{(mha)}$ is input into the Transformer network:
\begin{equation}
    \begin{aligned}
        \Wtwo'  & = \mathrm{LayerNorm}(Z+E_2^\mathrm{(mha)}),                                    \\
        \Wtwo'' & = \mathrm{Linear}\left(\mathrm{GELU}\left(\mathrm{Linear}(\Wtwo')\right)\right), \\
        \Wtwo   & = \mathrm{LayerNorm}\left(\Wtwo'+\Wtwo''\right),
        \label{eq:w2}
    \end{aligned}
\end{equation}
where $\mathrm{Linear}\left(\mathrm{GELU}\left(\mathrm{Linear}(\cdot)\right)\right)$ represents a multi-layer feedforward network, and $\mathrm{LayerNorm}(\cdot)$ denotes layer normalization.

\subsection{Periodic-wise Representation Learning (PRL)}
The network used to extract $\Wthree$ has the same structure as the network used to extract $\Wtwo$. When the feature dimensions input to the network differ, it is capable of capturing dependencies of different dimensions. To extract $\Wthree$, a transpose is first performed to move the subsequence dimension to the last axis, as illustrated in \cref{fig:g3} bottom. At the same time, the operations in \Cref{eq:e2,eq:qkv2,eq:w2} are carried out, which we denote concisely as:
\begin{equation}
    \Wthree = g_3\left((Z+\Wtwo)^\top\right)^\top.
\end{equation}
At this point, the network can effectively extract $\Wthree$. Subsequently, the linear decoder projects the learned representations onto the target prediction time domain:
\begin{equation}
    {\Xout_\mathrm{ds}} = \mathrm{Decoder}\left(\Wthree\right).
\end{equation}
It then combines them through inverse downsampling:
\begin{equation}
    \Xout = \mathrm{De{\text -}DownSample}\left({\Xout_\mathrm{ds}}\right),
\end{equation}
to form a predicted tensor of length $L_\mathrm{out}$, where $\Xout\in\mathbb{R}^{N_c\times L_\mathrm{out}}$ denotes the predicted results.
\renewcommand{\arraystretch}{0.7}
\begin{table}[t]
    \centering
    \caption{Ablation results of \OurModel}\label{tab:ablation_results}
    \vspace{-6pt}

    \setlength{\tabcolsep}{3pt}
    \begin{adjustbox}{max width=0.9\linewidth}
        \begin{threeparttable}
            {\sc
                \small
                \begin{tabular}{@{}llccccccc@{}}
                    \toprule
                    \multicolumn{2}{c}{Ablation} & Full & $\Wone$          & $\Wtwo$           & $\Wthree$         & $\Wone,\Wtwo$     & $\Wone,\Wthree$ & $\Wtwo,\Wthree$         \\
                    \midrule
                    \multirow{2}{*}{ETTh2}       & MSE  & \firstres{0.366} & \secondres{0.369} & 0.371             & 0.384             & 0.375           & 0.380           & 0.385 \\
                                                 & MAE  & \firstres{0.393} & \secondres{0.395} & 0.396             & 0.402             & 0.399           & 0.400           & 0.408 \\
                    \multirow{2}{*}{ETTm2}       & MSE  & \firstres{0.278} & \secondres{0.279} & 0.280             & 0.281             & 0.285           & 0.285           & 0.287 \\
                                                 & MAE  & \firstres{0.318} & \secondres{0.320} & \secondres{0.320} & \secondres{0.320} & 0.327           & 0.324           & 0.324 \\
                    \multirow{2}{*}{ECL}         & MSE  & \firstres{0.163} & 0.190             & \secondres{0.170} & 0.176             & 0.201           & 0.207           & 0.190 \\
                                                 & MAE  & \firstres{0.252} & 0.270             & \secondres{0.258} & 0.265             & 0.278           & 0.284           & 0.277 \\
                    \multirow{2}{*}{Solar}       & MSE  & \firstres{0.189} & 0.221             & \secondres{0.193} & 0.197             & 0.265           & 0.264           & 0.268 \\
                                                 & MAE  & \firstres{0.241} & 0.260             & \secondres{0.243} & 0.254             & 0.320           & 0.310           & 0.350 \\
                    \multirow{2}{*}{Traffic}     & MSE  & \firstres{0.431} & 0.454             & 0.462             & \secondres{0.448} & 0.509           & 0.510           & 0.530 \\
                                                 & MAE  & \firstres{0.268} & \secondres{0.278} & 0.282             & 0.292             & 0.306           & 0.321           & 0.334 \\
                    \bottomrule
                \end{tabular}}
            \begin{tablenotes}
                \small
                \item The results are averaged over four prediction horizons.
            \end{tablenotes}
        \end{threeparttable}
    \end{adjustbox}
    \vspace{-16pt}
\end{table}

\section{Experiment}\label{sec:experiment}

The performance of the proposed \OurModel is evaluated on eight widely used benchmark datasets, including ETT (ETTh1, ETTh2, ETTm1, ETTm2), Weather, Electricity (ECL), Traffic, and Solar. The data preprocessing procedure follows prior works~\cite{lin-2025-TQNet}, including the instance normalization strategy. All experiments were implemented using the PyTorch framework and executed on an NVIDIA RTX 4090 GPU equipped with 24 GB of memory. The training objective used an L1 loss function optimized with the Adam optimizer~\cite{Kingma-2014-AdamAM}.

\subsection{Long-term Forecasting}
\textbf{Setups.} We evaluated the performance of \OurModel against seven mainstream methods, including TQNet~\cite{lin-2025-TQNet}, Twinsformer~\cite{zhou-2025-twinsformer}, Leddam~\cite{yu-2024-Leddam}, iTransformer~\cite{liu-2024-itransformer}, PatchTST~\cite{nie-2023-patchtst}, Crossformer~\cite{zhang-2023-crossformer}, and FEDformer~\cite{zhou-2022-fedformer}. For long-term forecasting, following prior works~\cite{lin-2025-TQNet,zhou-2021-informer,wu-2021-autoformer,liu-2024-itransformer}, we set the prediction horizons to $L_\mathrm{out}\in\{96,192,336,720\}$, while fixing the look-back window length to $L_\mathrm{in}=96$. Mean Squared Error (MSE) and Mean Absolute Error (MAE) were employed as evaluation metrics. The best results are highlighted in \firstres{bold}, while the second-best results are \secondres{underlined}.

\textbf{Result Analysis.} As shown in \Cref{tab:full_results}, \OurModel consistently achieves state-of-the-art performance, obtaining the lowest average MSE and MAE across all datasets and prediction horizons. Specifically, compared with TQNet~\cite{lin-2025-TQNet} and Crossformer~\cite{zhang-2023-crossformer}, \OurModel reduces MSE/MAE by 2.22\%/2.45\% and 19.37\%/19.29\%, respectively, on average. On datasets with pronounced periodic variations (ETTh1, ETTh2, and Solar), \OurModel outperforms the runner-up TQNet, reducing MSE/MAE by 4.76\%/2.53\%, 3.17\%/2.24\%, and 4.55\%/5.86\%, respectively. Overall, \OurModel ranks within the top two in 58 out of 64 evaluation metrics, including 54 first-place results.

\textbf{Complexity Analysis.}
\OurModel balances computational efficiency and prediction accuracy by leveraging downsampling to reduce sequence length and adopting Efficient Attention to lower complexity from $O(n^2)$ to $O(n)$. These optimizations make \OurModel a robust choice for long-term time series forecasting.

\subsection{Ablation Studies}

\subsubsection{Ablation on $\Wone$, $\Wtwo$, and $\Wthree$}
As shown in \Cref{tab:ablation_results}, removing $\Wone$, $\Wtwo$, or $\Wthree$ individually results in a significant accuracy loss. For instance, on the ETTh2 and ETTm2 datasets, the highest accuracy loss occurs when $\Wthree$ is removed, highlighting the importance of inter-subsequence dependencies in datasets with clear periodic variations. Conversely, removing $\Wone$ has the least impact on ETT datasets, which have only seven channels and relatively weak inter-channel dependencies. However, in datasets with a large number of channels, such as ECL ($N_c=321$), Solar ($N_c=137$), and Traffic ($N_c=862$), removing $\Wone$ leads to a more significant accuracy loss, verifying the effectiveness of inter-channel dependencies.

Furthermore, removing any two of $\Wone$, $\Wtwo$, and $\Wthree$ simultaneously leads to a greater accuracy drop, with the largest reduction observed when $\Wtwo$ and $\Wthree$ are removed together. This highlights the necessity of modeling both intra-subsequence and inter-subsequence dependencies, as the ablated models consistently underperform compared to the full model.

\renewcommand{\arraystretch}{0.7}
\begin{table}[t]
    \centering
    \caption{Performance of different interval on ETTh1}\label{tab:different_period}
    \vspace{-6pt}

    \setlength{\tabcolsep}{3pt}
    \begin{adjustbox}{max width=\linewidth}
        \begin{threeparttable}
            {\sc
                \begin{tabular}{@{}cccccccccccccccc@{}}
                    \toprule
                    Interval & \multicolumn{2}{c}{2} & \multicolumn{2}{c}{4} & \multicolumn{2}{c}{8} & \multicolumn{2}{c}{16} & \multicolumn{2}{c}{24}                                                                           \\
                    Horizon                     & MSE                   & MAE                   & MSE                   & MAE                    & MSE                    & MAE               & MSE   & MAE   & MSE              & MAE              \\
                    \midrule

                    96                        & 0.373                 & 0.392                 & 0.373                 & 0.392                  & \secondres{0.370}      & \secondres{0.389} & 0.386 & 0.398 & \firstres{0.363} & \firstres{0.388} \\
                    192                       & 0.425                 & 0.420                 & \secondres{0.422}     & \secondres{0.420}      & 0.431                  & 0.425             & 0.436 & 0.428 & \firstres{0.414} & \firstres{0.416} \\
                    336                       & 0.464                 & 0.438                 & \secondres{0.460}     & \secondres{0.437}      & 0.494                  & 0.462             & 0.482 & 0.456 & \firstres{0.453} & \firstres{0.435} \\
                    720                       & \secondres{0.465}     & \secondres{0.458 }    & \secondres{0.465}     & 0.461                  & 0.553                  & 0.509             & 0.480 & 0.473 & \firstres{0.451} & \firstres{0.451} \\
                    \bottomrule
                \end{tabular}}
        \end{threeparttable}
    \end{adjustbox}
    \vspace{-17pt}
\end{table}

\subsubsection{Impact of Downsampling Interval}
This section examines the impact of setting incorrect downsampling intervals on model performance. We conducted ablation experiments on the ETTh1 dataset using different fixed periods (2, 4, 8, 16, 24). Notably, the minimum period calculated via the autocorrelation function (ACF) is 24 (daily period). As shown in \Cref{tab:different_period}, when the downsampling interval is set to 24, the model achieves the best performance, demonstrating that aligning the downsampling interval with the data's inherent periodicity enhances the capture of periodic patterns. Additionally, smaller downsampling intervals (e.g., 2 or 4) also yield competitive results, likely due to their ability to capture finer-grained trends in the time series.

\section{Conclusion}
In this paper, we tackled the challenge of capturing complex dependencies in time series forecasting.
Unlike existing approaches that primarily focus on temporal dependencies within downsampled subsequences, we emphasized the importance of inter-subsequence dependencies.
To address this, we proposed \OurModel, a novel framework that simultaneously models inter-channel, intra-subsequence, and inter-subsequence dependencies, enabling a more comprehensive representation of time series data.
Extensive experiments conducted on multiple real-world datasets demonstrated the effectiveness and robustness of our method, establishing new state-of-the-art performance in long-term time series forecasting.

%
\balance
\printbibliography

\end{document}